A possible gut microbiota basis for weight gain side effects of antipsychotic drugs

Harshad Joshi[1,2], Ankita Parihar[1], Dazhi Jiao[1], Shwetha Murali[1], David J. Wild[1]

[1]School of Informatics and Computing, Indiana University, Bloomington, Indiana, USA

[2]Department of Chemistry, Indiana University, Bloomington, Indiana, USA

Email addresses:

    AP: aparihar@indiana.edu

    HJ: hjoshi@indiana.edu

    DJ: djiao@indiana.edu

    SM: smurali@indiana.edu

    LB: lbuehner@indiana.edu

    DW: djwild@indiana.edu

**Abstract**

Weight gain is a well-established side effect of both conventional and newer anti-psychotic drugs, but the cause is not well understood. Recent studies correlate obesity with the presence or absence of particular genetic sequences in the gut microbiota. We identified strong associations between protein targets of antipsychotics and microbiota sequences directly related to weight regulation in human body, leading to a potential metagenomic mechanism of action. Further experimental study is recommended.

**Introduction**

First generation antipsychotic drugs (FGAs – butyrophenones, phenothiazenes such as chlorpromazine, and thioxanthenes) are associated with a variety of side effects including weight gain[1-3] and related conditions such as non–insulin-dependent diabetes mellitus[4]. Hypotheses on the role of antipsychotics in regulating weight of patients is widely documented[5,6]. A study in 1988 of over 200 patients[7] showed a fourfold increase in weight-gain as a side effect of receiving antipsychotics at least four times that in the general population. Second generation antipsychotics (SGAs) also show equal or higher weight gain than FGAs, with Clozapine and Olzapine the most highly implicated[8-10] including in children and adolescents[11]. Amongst other SGAs, risperidone and quetiapine appear to have intermediate effects where as aripiprozole and ziprasidone are associated with little or no significant weight gain. Studies on the risks and the nature of side-effects of both FGAs and SGAs have shown substantial link with obesity, but the mechanisms responsible are still unidentified[12].

It is known that presence or absence of certain gut microbial communities regulate total body fat of the host, as they affect the amount of energy extracted from the diet[13-18]. Studies on lean and obese mice and humans suggest that the gut microbiota influences the efficiency of calorie harvest from diet and further the energy usage and storage. The studies have linked obesity with two dominant bacterial divisions in gut micro-biota: bacteroidetes and firmicutess. A study was carried out on 154 individuals to study the fecal microbial communities of adult female monozygotic and dizygotic twin pairs concordant for leanness or obesity, and their mothers. The study identified a set of obesity related protein sequences (differential genes) in human gut microbe, which participate in mechanisms related to the development of fat, insulin resistance and low-grade inflammation that characterize obesity. V2 and V6 16S rRNA gene sequencing datasets and 21 obesity-associated protein sequences from *E.rectale, E.eligens, B.theta3731, B.theta7330, and B.WH2* were released (dataset from this study)[19]. Gut microbes can cause variation in the toxicity and efficacy of drugs in individuals and affect absorption and bioavailability of drugs in the host, and thus metagenomic considerations are likely to be important in developing more effective drugs and in future personalized health-care paradigms. Over the last two decades the dietary–microbe–host interactions and role of gut microbes affecting physiological processes regulating body weight, have received intense investigation [23-25]. Further, novel methods to manipulate commensal microbial composition through combinations of antibiotics, probiotics and prebiotics are being tested for therapeutic approach[26-27].

**Materials and Methods**

In this work we sought to examine whether weight regulation, as a side effect could be explained by potential drug action on microbial genes, which are known to be differentially associated with obesity[19]. Since there is very little published data on the action of drugs on microbial genes, our approach involved identifying human genes with high sequence similarity to these differential genes, and examining the bioactivity profiles of drugs against these human genes (with the assumption that drugs with known bioactivity against the highly similar human genes have a high chance of activity against the microbial counterparts). Sequence data on 3,19,812 proteins from 89 frequent microbial genomes was extracted from the KEGG database[28-30], as a reference set and was added to sequence data for the 21 differential obesity genes identified in the obesity study[19]. We carried out a sequence similarity search on this combined set using BLASTP[31] against sequences for enzyme and drug targets in humans from DrugBank[32] with criteria[33] for a match being significance  E-value <10e-2 and percent identity > 47. 741 matching human genes were found, 40 matching differential microbial genes and 701 matching the reference set genes. We used approved drugs dataset from DrugBank database to extract the drugs known to bind to these genes [34-35]. We identified 145 and 611 FDA approved drugs targeting our set of the differential and non-differential matching genes respectively. Finally, we profiled side-effects of these selected drugs using the SIDER database[36].

**Results and Discussion**

The identified drugs were categorized based on DrugBank annotations and the frequency of (occurrence) drug categories related to differential and non-differential gut microbial gene list were shortlisted. Out of 9 drugs with weight gain as a side effect, 7 were of the antipsychotic class. The antispychotics with weight gain side effects showed a significant ($p<0.001$) association with the differential genes when compared with other drugs (mean 6.625 associations versus 1.229), with the one antispsychotic without a registered weight gain side effect (fencamfamine) having just 3 differential gene associations. Interestingly, the antispychotics also show a significantly lower ($p<0.001$) association with non-differential genes than the other drugs (mean 5.75 associations versus 16.772). Table 1 shows a list of the antipsychotic drugs with associated differential human genes (i.e. those with high sequence similarity to the bacterial genes with weight-gain associations), organized by bacterium of the bacterial genes. All of the associated genes are from *Bacteriodes thetaiotaomicron* and *Eubacterium rectale*. Both of these have an important influence on host energy balance and there are indications of a direct link with body weight regulation, although the pathways involved are still under investigation[37-40]. The gene functions cover metabolite biosynthesis, transport, catabolism, metabolism and cellular metabolism.

**Conclusion**

We thus propose a plausible metagenomic mechanism of action for the weight regulation side effects of antipsychotics through their potential interruption of pathways in the gut *Bacteriodes thetaiotaomicron* and *Eubacterium rectale* bacteria. We believe further investigation is warranted, including direct testing of the effects of the drugs on these bacteria, examination of weight regulation side-effects of these drugs on subpopulations of patients who have different gut states[41] or ethnic populations with gut bacterial diversity, and on the association of these bacteria with weight regulation in the body. Depending on the outcome of these studies, probiotic supplement of these bacteria in patients taking antipsychotics may be evaluated. The application of this methodology to other outcomes, including cancers, is also being evaluated.

Author contributions

AP, HJ, DJ, SM, LB carried out the work

DW, AP and HJ conceived the ideas and wrote the paper

Competing financial interests statement

The authors have no competing financial interests.

Table

| Zipradisone | *Bacteriodes thetaiotaomicron* {CYP1A2, CYP2R1, CYP17A1, PGD}; |
| --- | --- |
| Chlorpromazine | *Bacteriodes thetaiotaomicron* {Steryl-sulfatase, SLC35D1, cyaC, CYP17A1, CYP17A2, CYP2R1, pph}; *Eubacterium rectale* {pph, fixL, phoQ, envZ, TM_1359, fixL} |
| Palonosetron | *Bacteriodes thetaiotaomicron* {SLC35D1, cyaC, CYP2R1, PRODH}; *Eubacterium rectale* {pph, fixL, cheA, cyaC, ABCA1, CYP2E1} |
| Clozapine | *Bacteriodes thetaiotaomicron* {CYP17A1, PGD, phoQ} |
| Promazine | *Bacteriodes thetaiotaomicron* {cyaC, CYP2R1, PRODH, CYP17A1} |
| Thioridazine | *Bacteriodes thetaiotaomicron* {SLC35D1, cyaC, CYP2R1, CYP17A1, PGD, CYP1A2}; *Eubacterium rectale* {CYP2E1, envZ, TM_1359} |
| Olanzapine | *Bacteriodes thetaiotaomicron* {SLC35D1, cyaC, PRODH, CYP17A1, PGD, pph, CYP1A2} |

*Table 1. Antipsychotic drugs with human genes which (i) are binding targets for the drugs and (ii) have high sequence similarity to gut bacteria genes differentially associated with weight gain. Genes are organized by the parent bacterium of the bacterial genes.*